\documentclass[journal=jpccck,manuscript=article,layout=traditional]{achemso}

\usepackage{graphicx,epsfig,multirow}

\author{Ersen Mete}\email{emete@balikesir.edu.tr}
\affiliation{Department of Physics, Bal{\i}kesir University, Bal{\i}kesir 10145,
Turkey}
\author{Ay\c{s}en Y{\i}lmaz}
\affiliation{Department of Chemistry, Middle East Technical University, Ankara
06800, Turkey}
\author{Mehmet Fatih Dan{\i}\c{s}man}\email{danisman@metu.edu.tr}
\affiliation{Department of Chemistry, Middle East Technical University, Ankara
06800, Turkey}

\title[Carboranethiol Self-Assembled Monolayer on Au(111)]
{A Density Functional Theory Investigation of Carboranethiol Self-Assembled 
Monolayer on Au(111)}

\keywords{DFT Calculations, Carboranethiol, Self-Assembled Layers, Au Surface}

\begin{document}

\begin{abstract}
Isolated and full monolayer adsorption of various carboranethiol 
(C$_2$B$_{10}$H$_{12}$S) isomers on gold (111) surface have been investigated 
using both the standard and van der Waals density functional theoretical 
calculations. The effect of differing molecular dipole moment orientations on 
the low energy adlayer geometries, the binding characteristics and the 
electronic properties of the self-assembled monolayers of these isomers 
have been studied. Specifically, the binding energy and work 
function changes associated with different molecules show a correlation with 
their dipole moments. The adsorption is favored for the isomers with dipole 
moments parallel to the surface. Of the two possible unit cell structures, the 
(5$\times$5) was found to be more stable than the 
($\sqrt{19}\times\sqrt{19}$)R23.4$^\circ$ one.
\end{abstract}

\section{Introduction}
Thiol self-assembled monolayers (SAMs) on metal surfaces are ubiquitous systems 
due to their easy adaptability in many different applications ranging from 
bio-sensors to electronics.\cite{Frasconi,Love,Schreiber,Vericat} The 
utility of these systems is that by designing/using an appropriate thiol 
molecule, the properties of the metal surfaces can be altered in a controlled 
way. The properties of thiol SAMs are governed by the balance between the 
intermolecular and molecule-surface interactions and have been the subject of 
intense experimental\cite{Love,Schreiber,Vericat,Albayrak1,Albayrak2,Albayrak3}
and computational\cite{Gronbeck1,Ferrighi,Hayashi,Gronbeck2,Gronbeck3,
Yourdshahyan1,Yourdshahyan2,Vargas,Molina,Otalvaro,Rusu,Fertitta,AbuHusein,
Verwuester,Fajin,Lustemberg,YWang,Torres1,Torres2,Longo,Cossaro,Mazzarello,
Rousseau,DeRenzi1,ZZhang,Fischer,Osella,Barmparis,JGWang1,JGWang2,Sun,DeRenzi2} 
studies. By playing with the chemical nature of the thiol molecules this balance 
can be altered and SAMs with different structural, interface and/or surface 
properties could be obtained. 

One very important property, especially for electronic applications, is the 
work function of the metal surface coated with the SAM which can be controlled 
by tuning the dipole moment of the molecules forming the SAM.\cite{Rusu,
AbuHusein,Verwuester,Rousseau,DeRenzi1,Osella,Heimel,Campbell} To alter the work 
function either single component SAMs made up of thiol molecules with 
appropriate dipole moment or mixed SAMs made up of two different thiol molecules 
each with a different dipole moment can be employed.\cite{Otalvaro,Campbell,
Chen,Venkataraman,Alloway,Wu,Xu} In both cases, most of the time, the dipole 
moment of the molecules are tuned by adding a functional group to the backbone 
or to the end of the molecules. This approach however not only changes the 
dipole moment but also alters the geometry of the molecule which in turn may 
result in SAMs with different structural properties which is coupled to the 
electronic properties of the film/surface. To be able to study the effect of 
geometry and the chemical/electronic nature of the molecule on the properties of 
the film independently, hence, it is necessary to decouple these two changes in 
the molecule. Dicarbacloso-dodecaborane thiols (C$_2$B$_{10}$H$_{12}$S, will be 
referred to as carboranethiols) is an outstanding alternative to this end, since 
by playing with the positions of the carbon and sulfur atoms,  the electronic 
properties of the molecule (i.e. dipole moment)  can be altered without changing 
geometry. This fact combined with their chemical stability and almost spherical 
shape make CTs unique molecules for studying the fundamental properties of thiol 
SAMs and for preparing films that can be used in many different applications. 
Hence, SAMs of CTs and their derivatives have attracted increasing interest in 
recent years.\cite{Lubben,Base1,Base2,Base3,Scholz,Wrochem,Kim,Hohman1,Hohman2,
Bould,Thomas1,Thomas2} 

\begin{figure}
\includegraphics[width=8.3cm]{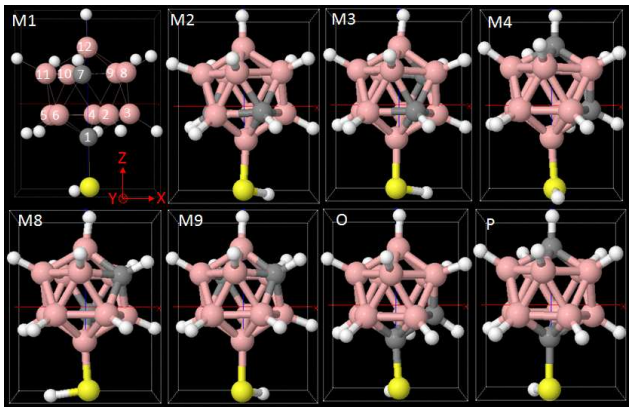}
\caption{The vdW-optimized structures of CT isomers. The numbering 
scheme is given in the M1 molecule.\label{fig1}}
\end{figure}

Recently, L\"{u}bben and coworkers have demonstrated that by using pure and 
mixed SAMs of two carborane dithiol isomers 
[1,2-(HS$_2$)-1,2-C$_2$B$_{10}$H$_{10}$ and 
9,12-(HS$_2$)-1,2-C$_2$B$_{10}$H$_{10}$] with opposite dipole moments, the 
surface potential of silver surfaces could be tuned. Weiss et al., on the other 
hand, used two CT isomers [1-HS-1,7-C$_2$B$_{10}$H$_{11}$ (M1) and 
9-HS-1,7-C$_2$B$_{10}$H$_{11}$ (M9), see figure 1 for naming of CT isomers]  and 
were able to tune the  gold work function over a range of 0.8 eV by preparing 
mixed SAMs of these two isomers in different ratios.\cite{Kim} Then, they used 
such SAM coated gold surfaces as electrodes of organic field effect transistors 
and observed an improvement in the device characteristics. In addition they 
found that M1 adsorbs on the Au(111) surface stronger than M9, based on contact 
angle and reflection absorption IR spectroscopy results, and attributed this to 
dipole-dipole interactions of the molecules in the SAM. In case of M1 the dipole 
moment vector of the molecule is parallel to the gold surface (see 
Figures~\ref{fig1} and \ref{fig2}) whereas for M9 the direction is 
perpendicular to the surface. Hence Weiss \textit{et al.} proposed that the 
head-to-tail orientation of the dipoles of the M1 molecules could yield a more 
favorable adsorption when compared with M9. In addition, very recently they 
showed that dipole-dipole interactions to be highly defect tolerant based on 
STM measurements\cite{Thomas1} which can be interpreted as another outcome of 
the importance of the dipole-dipole interactions on determining thiol film 
structure and adsorption strength. Finally, based on scanning tunneling 
microscope images, they suggested two possible unit cell [(5 $\times$ 5) and 
($\sqrt{19} \times \sqrt{19}$)R23.4$^o$] structures for M1 and M9 SAMs which 
are depicted in Figure~\ref{fig3}.\cite{Hohman2}

Inspired by these experimental results we set out to perform a computational 
study to investigate the relation between the dipole moments of the molecules 
and their binding strengths and monolayer electronic and crystal structures. To 
this end we have calculated the dipole moments of all possible CT isomers shown 
in Figure~\ref{fig1} by using density functional theory. Then we have modeled 
isolated molecules on Au(111) surface and determined the binding energies. 
Finally for the isomers M1 and M9 we have performed calculations for different 
full monolayer structures and determined the preferred crystal structure and 
the effect of the crystal structure on the electronic properties of the 
interface.

\section{Theoretical Methods}

In order to describe the atomic and electronic structures of isolated and full 
monolayer adsorption of carboranethiols on the flat gold surface, periodic 
boundary density functional theory (DFT) calculations were carried out based on 
the projector augmented-wave (PAW)\cite{Blochl} method as implemented 
in VASP.\cite{Kresse1,Kresse2,Kresse3} Single particle valence states were 
expanded in terms of plane waves up to a kinetic energy cutoff of 400 eV. The 
electron-electron exchange and correlation interactions have been included using 
both the standard Perdew-Burke-Ernzerhof (PBE) functional within the 
generalized gradient approximation (GGA) and the dispersion corrected 
optB86b-vdW functional\cite{Klimes} within the van der Waals density functional 
(vdW-DF) approach.\cite{Dion} The vdW correction mimics the attractive 
electronic interactions beyond the equilibrium separations. Therefore, the basic 
idea is to include the long range part of the correlation energy as a fully 
nonlocal functional of the charge density.

Before moving on to the surface properties, we tested PBE and optB86b-vdW 
(exchange-correlation) XC functionals against the lattice parameter of the  
bulk gold (fcc). The standard PBE gives a value of 4.160 {\AA}. The vdW 
correction leads to 4.125 {\AA} which better compares with the experimental 
value of 4.078 {\AA}.\cite{Wyckhoff} We modeled the flat gold surface as a four 
atomic layer slab in a supercell which also contains a vacuum region with a 
height of 21 {\AA} along the surface normal. This vacuum separation reduces to 
14 {\AA} after CT molecule adsorption on the surface. Initial coordinates of 
gold atoms were taken from their bulk positions. Then, we performed a full 
optimization of the ionic coordinates and lattice translation vectors based on 
the variational minimization of the Hellmann-Feynman forces by requiring each 
spatial component to be less than 0.01 eV/{\AA} on each atom. The average 
nearest neighbor Au-Au bonds were found as 2.917 {\AA} with and as 2.942 {\AA} 
without the dispersive corrections. The standard XC functionals tend to 
overestimate the bond lengths especially in metallic systems. This is basically 
due to the local density approximation (LDA). It tries to parametrize the XC 
energy as a functional of the electron density from the uniform distribution of 
an homogeneous electron gas with the same charge density.

For calculations involving molecular adsorbates on the surface, the gold atoms 
at the bottom two layers were frozen to their fully relaxed positions.
The ($5\times 5$) surface unit cell is considered for the adsorption of an 
isolated CT on Au(111). This supercell is large enough to put at least 9.9 
{\AA} separation between the periodic images of the CT adsorbates on the 
surface.

In the full monolayer CT coverage on flat gold, we considered the ($5\times 5$) 
and ($\sqrt{19}\times\sqrt{19}$)R23.4$^\circ$ structures as a probable surface 
phases as suggested by recent experiments.\cite{Hohman2} The surface Brillouin 
zone integrations were carried out over $k$-point samplings with 
$\Gamma$-centered 5$\times$5$\times$1 meshes. Our tests with denser 
$k$-point grids showed that the total energies were converged to an accuracy 
of 10 meV. The density of states (DOS) calculations were performed
with 7$\times$7$\times$1 $k$-point mesh using the tetrahedron method. 

The dissociative adsorption energies of Au(111) surfaces with isolated and full 
monolayer CTs can be calculated by,
\[
E_{\rm ads}=E_{{\rm CT}/{\rm Au(111)}}-E_{\rm Au(111)}-n(E_{\rm 
CT}-\frac{1}{2}E_{H_2}),
\]
where $E_{{\rm CT-H}/{\rm Au(111)}}$, $E_{\rm Au(111)}$, $E_{\rm CT}$ and 
$E_{H_2}$ are the total energies of the Au(111) slab with CTs which lost their 
tail hydrogens, of the clean Au(111) slab, of a single CT in a big box and of a 
hydrogen molecule in vacuum, respectively. In this expression, $n$ is the 
number of CT adsorbates on the surface.

\begin{table}[htb]
\caption{Relative total energies, $E_{\rm t}$, and dipole moments, 
$\mu$, of carboranethiol (CT) compounds in the gas phase 
calculated using both PBE and optB86b-vdW functionals.\label{table1}}
\begin{tabular}{lccccccccccc}\hline \\[-4mm]
\multirow{2}{*}{CT} & \multicolumn{5}{c}{PBE} && \multicolumn{5}{c}{vdW-DF} \\ 
\cline{2-6} \cline{8-12} \\[-4mm]
& $E_{\rm t}$ & $\mu_x$ & $\mu_y$ & $\mu_z$ & $\mu$ && 
$E_{\rm t}$ & $\mu_x$ & $\mu_y$ & $\mu_z$ & $\mu$ \\[1mm]\hline
M1 & 0.999 &-0.56 &-1.55 &-0.66 & 1.78 && 1.008 &-0.58 &-1.53 &-0.64 & 1.76 \\
M2 & 0.002 &-0.41 &-0.43 &-1.53 & 1.64 && 0.013 &-0.40 &-0.43 &-1.50 & 1.61 \\
M3 & 0.020 & 3.08 & 0.05 & 0.81 & 3.18 && 0.040 & 3.05 & 0.05 & 0.77 & 3.15 \\
M4 & 0.000 & 2.16 &-0.40 & 2.29 & 3.17 && 0.000 & 2.13 &-0.39 & 2.26 & 3.13 \\
M8 & 0.008 & 1.59 & 1.21 & 0.84 & 2.16 && 0.018 & 1.58 & 1.20 & 0.82 & 2.15 \\
M9 & 0.065 & 1.27 & 1.53 & 3.34 & 3.89 && 0.085 & 1.26 & 1.52 & 3.33 & 3.87 \\
Ortho & 1.576 & 1.78 & 0.96 &-3.10 & 3.70 && 1.596 & 1.76 & 0.92 &-3.08&3.66\\
Para & 0.869 &-0.25 & 0.59 & 0.84 & 1.06 && 0.887 &-0.25 & 0.57 & 0.81 &1.02 \\ 
\hline
\end{tabular}
\end{table}

\section{Results and Discussion}

The gas phase structures of various CT variants were obtained with both 
the standard PBE and the modern vdW-DF functionals. The positional labeling of 
the carboranethiol isomers follow as shown in Figure~\ref{fig1}. Although 
the cage geometries are essentially similar their relative total energies and 
dipole moments are different (see Table~\ref{table1}).

\begin{figure*}
\includegraphics[width=16cm]{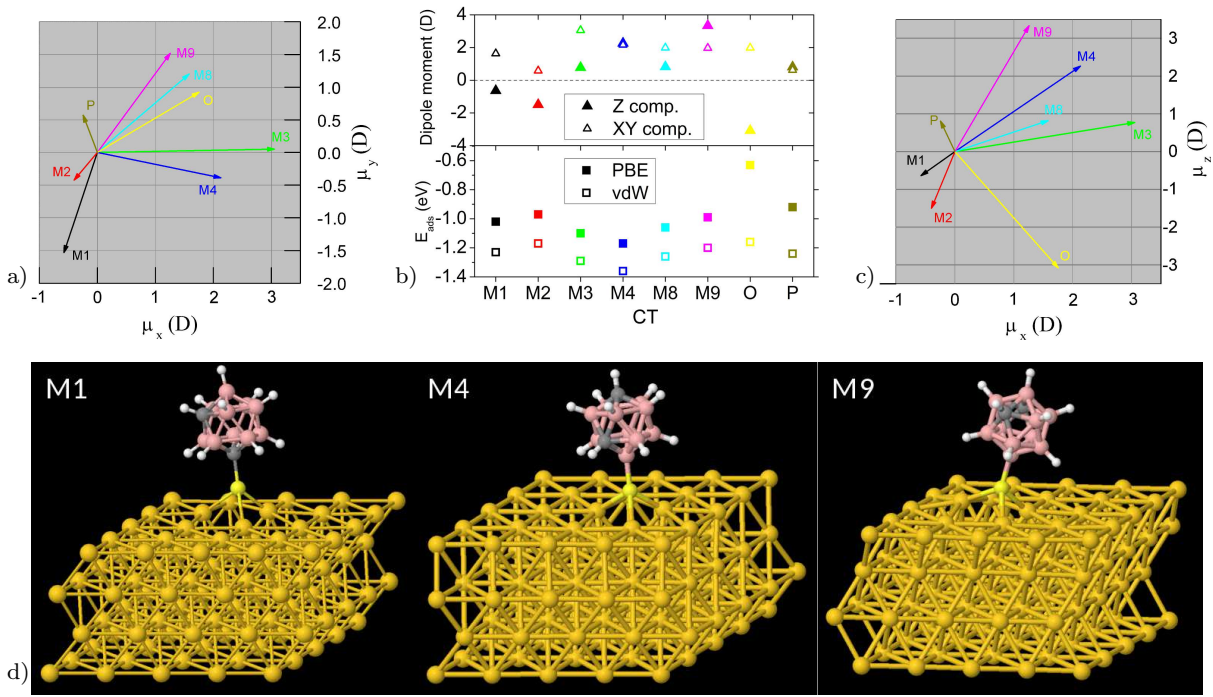} 
\caption{Plots showing the a) XY, c) XZ components of the dipole moments of 
the CT isomers in the gas phase. b) Chart showing the Z, XY 
components of the dipole moments and corresponding adsorption energies of CT 
isomers on Au(111). d) Optimized geometries of the isolated M1, 
M4 and M9 CTs on the gold surface. 
\label{fig2}}
\end{figure*}
 
\begin{figure*}[htb]
\epsfig{width=16cm,file=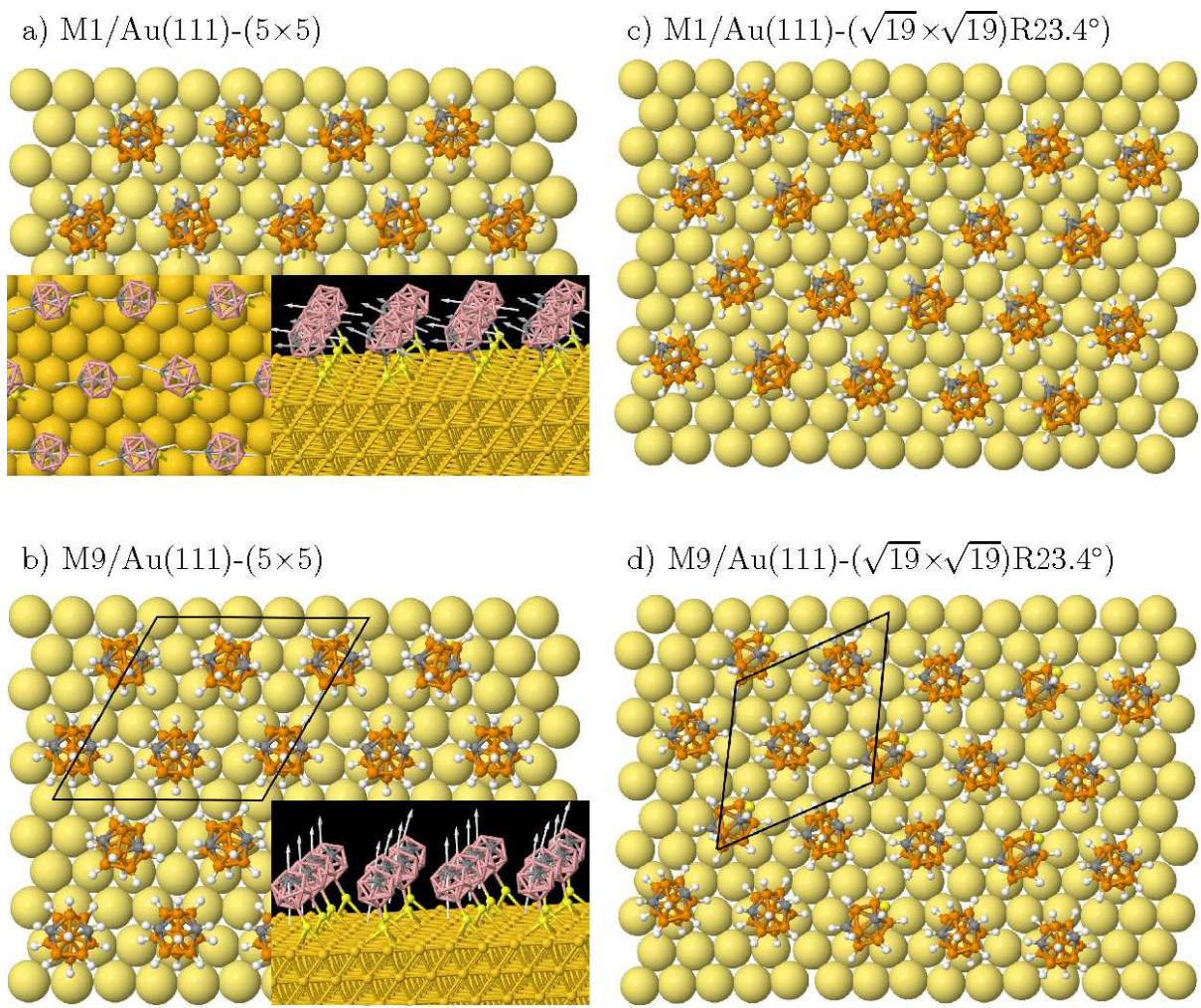} 
\caption{The adsorption geometry of ($5\times 5$) and 
($\sqrt{19}\times\!\sqrt{19}$)R23.4$^\circ$ structures of 1ML M1 and M9 on the 
unreconstructed Au(111) surface optimized using vdW-DF.\label{fig3}}
\end{figure*}

We considered all probable adsorption configurations of CTs on the 
unreconstructed gold (111) surface. For instance, the adsorption of M9 on gold 
at the bridge site is energetically 0.52 eV  higher than the most favorable 
position at the hollow site. In addition, when initially placed on top of one 
of the surface gold atoms, the molecule relaxes into the bridge configuration. 
We also checked and saw that a geometry optimization of an isolated CT starting 
from the bridge does not end up with the hollow position. This result is not 
enough to rule out the possibility of the coexistence of both adsorption 
types in an experimental realization. Indeed, our 1 ML results show that
while most of the molecules are adsorbed at the hollow site, small number of 
them come close to the bridge site as shown in Figure~\ref{fig3}.

\begin{table}[htb]
\caption{Relative total cell energies, $E_{\rm t}$, and dissociative adsorption 
energies, $E_{\rm ads}$, of Au(111) with a single isolated CT, calculated using 
PBE and optB86b-vdW methods.\label{table2}}
\begin{tabular}{lccccccccc}\hline\\[-4mm]
& \multicolumn{4}{c}{PBE} && \multicolumn{4}{c}{optB86b-vdW} 
\\ \cline{2-5}\cline{7-10} \\[-4mm]
CT & $E_{\rm t}$ & $E_{\rm ads}$ & $d_{S-Au}$ & $h$ && $E_{\rm t}$ & 
$E_{\rm ads}$ & $d_{S-Au}$ & $h$ 
\\[1mm]\hline\\[-4mm]
M1&1.14&-1.02&2.44, 2.47, 2.57&1.49&&1.15&-1.23&2.42, 2.44, 2.52&1.43\\
M2&0.21&-0.97&2.44, 2.46, 2.53&1.45&&0.20&-1.17&2.42, 2.44, 2.48&1.37\\
M3&0.10&-1.10&2.45, 2.45, 2.51&1.51&&0.10&-1.29&2.44, 2.45, 2.49&1.48\\
M4&0.00&-1.17&2.43, 2.45, 2.50&1.40&&0.00&-1.36&2.39, 2.41, 2.46&1.35\\
M8&0.13&-1.06&2.43, 2.45, 2.51&1.41&&0.11&-1.26&2.42, 2.43, 2.48&1.37\\
M9&0.24&-0.99&2.45, 2.46, 2.50&1.55&&0.19&-1.20&2.43, 2.45, 2.49&1.51\\
Ortho&2.11&-0.63&2.43, 2.44, 2.51&1.44&&1.79&-1.16&2.42, 2.43, 2.49&1.38\\
Para&1.12&-0.92&2.44, 2.47, 2.56&1.48&&1.00&-1.24&2.41, 2.42, 2.49&1.35\\ 
\hline
\end{tabular}
\end{table}

In the minimum energy adsorption, a single CT molecule attaches to the 
surface via its tail sulphur which shows three-fold coordination with the 
nearest neighbor gold atoms at the hollow site. The three S-Au bonds are not 
equal in length as presented in Table~\ref{table2}. For the isolated case of M9 
on Au(111), they are found as 2.45, 2.46, 2.50{\AA} with PBE and 2.43, 2.45, 
2.49 {\AA} with vdW-DF at the hollow position. Although the vdW-DF bond lengths 
are only slightly shorter than the PBE ones, the inclusion of the dispersive 
forces have non negligible effect on the minimum energy geometries especially 
on the surface gold atoms in the vicinity of the adsorption region. The 
standard PBE functional leads to considerable local distortion on the gold 
surface while it is less noticeable when vdW corrections apply. For instance, 
M9 sits on the hollow site above the surface gold triangle where the three 
Au-Au distances become 3.45, 3.49, 3.49 and 3.37, 3.37, 3.42 {\AA} with the PBE 
and optB86b-vdW functionals, respectively. 

In Figure~\ref{fig2}, the dipole moments and the adsorption geometries of the 
studied CT isomers are shown. When the z components of the dipole moments are 
considered, a correlation with the adsorption energies can be noticed. A dipole 
monent vector pointing above the surface results in lower binding energies with 
both PBE and optB86b-vdW functionals. Whereas dipole moment vectors pointing 
towards the surface increase binding energies.  

\begin{figure*}[htb]
\epsfig{width=14.5cm,file=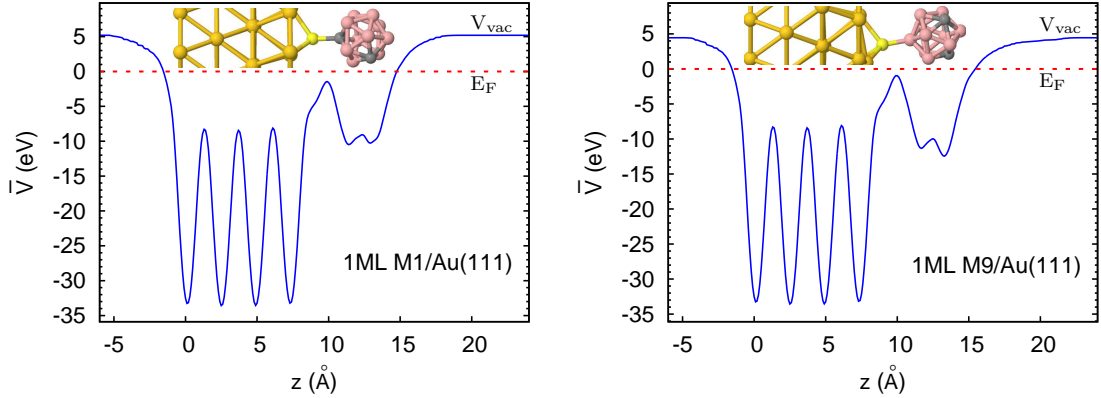} 
\caption{Electrostatic potential energy profiles of the Au(111) slab with M1 
and M9 CT adlayers. $\bar{\rm V}(z)$ is plane averaged along the 
surface normal. V$_{\rm vac}$ is the potential energy value at the 
vacuum regions and E$_{\rm F}$ is Fermi level.\label{fig4}}
\end{figure*}

The work function of the clean gold (111) surface is calculated using the 
expression $W=V(\infty)-E_{\rm F}$ as 5.15 eV. Here, $V(\infty)$ is the 
electrostatic potential in the vacuum and E$_F$ is the Fermi energy. We
obtained the real space electrostatic potential $V(x,y,z)$ in a 
self-consistent DFT calculation. Then the plane averaged potential
is given by
\begin{equation}
\bar{\rm V}(z)=\frac{1}{A}\,\int\hspace{-2mm}\int\limits_{\hspace{-3mm}\rm 
cell} dx\,dy\,\, V(x,y,z)  
\end{equation}
where A is the surface unit cell area. The plots for surfaces with M1 and M9 
SAM structures are presented in Figure~\ref{fig4}.

The dissociative adsorption of an isolated M9 molecule on the gold 
(5$\times$5) surface corresponding to 0.25 ML coverage yields a
work function of 4.81 eV. The same surface structure with an isolated 
M1 gives this value as 5.13 eV. For a full M9 adlayer on the gold surface, 
it becomes 4.45 eV while M1 SAM leads to a work function of 5.20 eV.
In other words, M9 SAM results in a reduction of 0.70 eV in the work
function of the gold surface while M1 SAM increases it by 0.05 eV.
When the electrostatic potential profiles of M1 and M9 SAM structures in 
Figure~\ref{fig4} are compared, the higher potential well depth of M1 
indicates a lower electron density in the molecular film relative to M9.
This implies a stronger interaction between the M1 layer and the gold surface.
Therefore, these results indicate a better binding in favor of M1 adlayer.
Weiss \textit{et al.}, in their experimental study, have reported a reduction 
of 0.4 eV for M9 and increase of 0.4 eV for M1 SAMs.\cite{Kim} Though 
the changes in the work function due to individual M1 and M9 SAMs we 
report here do not match very well with the experimental results, the total 
variation (0.75 eV) agrees very well with the experimental value (0.8 eV).

\begin{table}[htb]
\caption{Relative total cell energies, $E_{\rm t}$, the dissociative adsorption 
energies, $E_{\rm ads}$, and the heights of a full CT adlayer on 
Au(111) with two different phases calculated using the PBE and the optB86b-vdW 
exchange-correlation functionals.\label{table3}}
\begin{tabular}{lccccccccccccc}\hline\\[-4mm]
\multirow{3}{*}{CT} & \multicolumn{6}{c}{PBE} && 
\multicolumn{6}{c}{optB86b-vdW} 
\\ \cline{2-7}\cline{9-14} \\[-4mm]
& \multicolumn{3}{c}{(5$\times$5)} &
\multicolumn{3}{c}{($\sqrt{19}\!\times\!\!\sqrt{19}$)R23.4$^{\circ}$} && 
\multicolumn{3}{c}{(5$\times$5)} &
\multicolumn{3}{c}{($\sqrt{19}\!\times\!\!\sqrt{19}$)R23.4$^{\circ}$}
\\ \cline{2-7}\cline{9-14} \\[-4mm]
& $E_{\rm t}$ & $E_{\rm ads}$ & $h$ & $E_{\rm t}$ & $E_{\rm ads}$ & $h$ && 
$E_{\rm t}$ & $E_{\rm ads}$ & $h$ & $E_{\rm t}$ & $E_{\rm ads}$ & $h$ 
\\[1mm]\hline\\[-4mm]
M1 &3.55&-0.48&2.05&3.51&0.73&1.55&&2.74&-0.70&2.02&2.44&-0.35&1.49 \\
M9 &0.00&-0.30&2.05&0.00&0.79&1.56&&0.00&-0.51&2.01&0.00&-0.02&1.49 \\ \hline
\end{tabular}
\end{table}

\begin{figure}[htb]
\epsfig{width=8cm,file=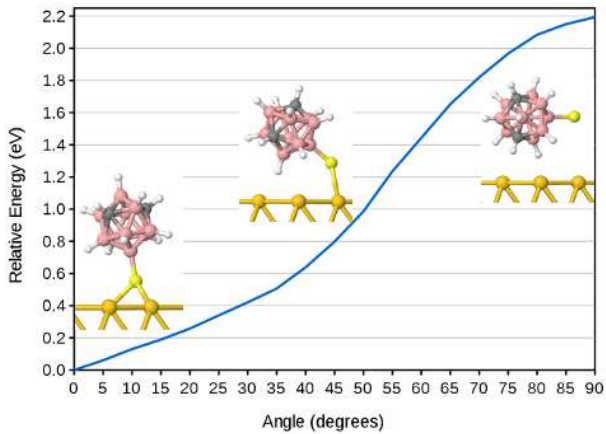} 
\caption{Relative energy of single M9 + Au(111) combined system with respect to 
the tilting angle of S-Au bond. The zero of the angle is set as the 
minimum energy vertical alignment of M9 that is not perfectly perpendicular 
to the surface.\label{fig5}}
\end{figure}

We have computed the relative cell energy of an isolated M9 as a function of 
the tilting angle as shown in Fig.~\ref{fig5}. M9 makes an angle of 
$\sim$10$^\circ$ with the surface normal at its minimum energy binding. The 
potential energy barrier is almost 2.1 eV between standing up and lying 
parallel orientations. This big energy difference can be attributed to the 
generation of strong Au-S bonds by M9 with large molecular dipole.   

\begin{figure*}[htb]
\epsfig{width=15cm,file=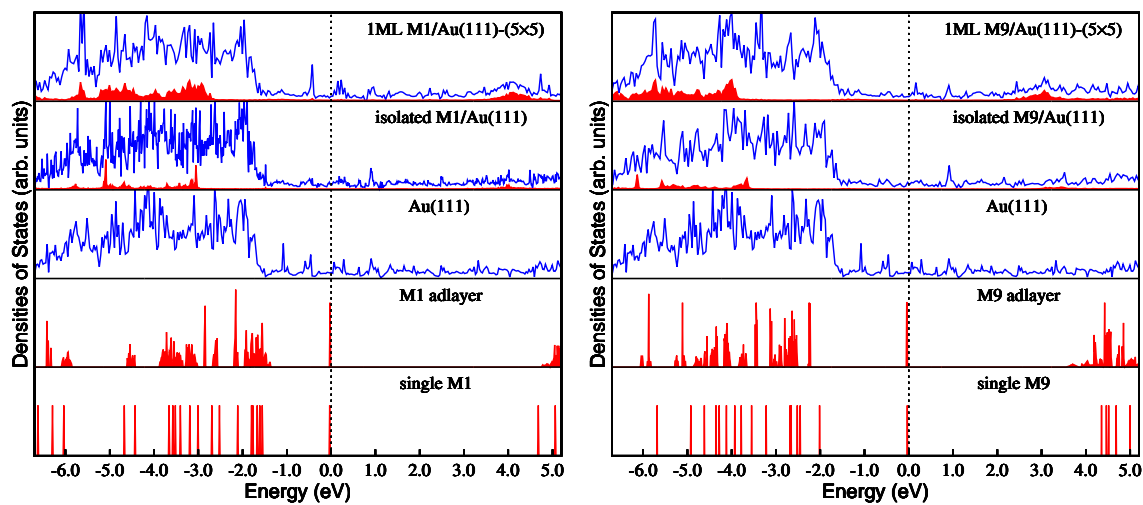} 
\caption{Calculated total and partial densities of states (DOS) of 
isolated and 1ML M1/M9 CT on Au(111). Shaded regions indicate the 
molecular contributions. The DOS plots of a single M1/M9 in the gas phase and 
of an adlayer without the gold slab are presented in the lower two 
panels, correspondingly.\label{fig6}}
\end{figure*}

The average value of the nearest neighbor spacings between the carboranethiol 
isomers at 1ML on Au(111)-(5$\times$5) are about 7.2 {\AA} and 7.0 {\AA} with 
and without vdW corrections. The corresponding mean values slightly change to 
7.4 {\AA} and 7.2 {\AA} on the ($\sqrt{19}\!\times\!\!\sqrt{19}$)R23.4$^\circ$ 
structure. These geometry optimization results are consistent with experimental 
value of 7.2$\pm$0.4 {\AA}.\cite{Hohman2} When the relative total energies of 
the flat gold surfaces with M1 and M9 adlayers are compared, the latter is found 
to be energetically more preferable than the former by 3.55 eV by PBE or 2.74 eV 
by vdW-DF per (5$\times$5) unit cell as seen in Table~\ref{table3}. On the 
other hand, the adsorption energy comparisons favor M1 over M9 in terms of 
stability at both low and full concentrations. The PBE functional appears to 
underestimate the adsorption energies. Especially, PBE does not give any 
binding for 1 ML CT on ($\sqrt{19}\!\times\!\!\sqrt{19}$)R23.4$^\circ$ unit 
cell yielding a positive adsorption energy. This result urges the 
importance of vdW corrections in the theoretical description of metal-organic 
systems. Hohman \textit{et al.} reported that M1 is the preferred species in 
mixed monolayers. Our vdW-DF results indicate a stronger adsorption 
in favor of M1 relative to M9 in good agreement with the 
experiments.\cite{Hohman2} The most striking difference of these similar CT 
isomers is their dipole moments. Therefore, the dipole-dipole interactions 
between the molecules has a significant role in the SAM stability on the gold 
surface.  

We have calculated the total and projected densities of states (DOS) of Au(111)
with M9 and M1 at low and full coverages using the vdW-DF theory calculations. 
We chose M9 and M1 because they cause work function changes in the opposite 
directions. We first stated with the electronic structures of a single M9(M1) 
in the gas phase and of an M9(M1) adlayer without the gold slab as shown in the 
lower two panels of Fig.~\ref{fig6}. The HOMO-LUMO separation of M9(M1) is 
found to be 4.4(4.7) eV. Due mostly to dipole-dipole interactions 
within the CT layers, the energy levels, in both cases, get slightly dispersed 
without showing a significant shift in their positions. The weakness of the 
coupling between the molecules show itself as a broadening of the energy 
levels mimicking a long-range correlation effect. The isolated and 1 ML M1 on 
Au(111) show similar characteristics. However, they differ from the M1 cases 
for which gold slab is absent. Due to the formation of strong Au-S bonds the 
DOS contributions of M1 species exhibit a significant red shift in the 
energy spectra of CT-Au(111) systems. Similar conclusions can be drawn for 
the M9 case. On the other hand, one can notice that M1 peak positions 
energetically lie lower relative to those of M9. Therefore, the 
binding characteristics of M1 is favored over M9. This result is also 
consistent with our calculated adsorption energies in Table~\ref{table3}.
Although the dipole-dipole interactions are weak in SAMs, the inclusion of vdW 
correction terms in the description of electronic structures of organic-metal 
interfaces becomes important. 

\section{Conclusions}

We carried out periodic density functional theory calculations by considering
isolated and full adlayer of carboranethiol isomers on the Au(111) surface.
In order to quantitatively show the effect of the long-range correlations 
on both the final geometries of CT-metal composite systems and their 
electronic structures, we included dispersive forces in a self-consistent 
implementation of the vdW-DF method. Then comparisons were made with 
the results obtained using the standard exchange-correlation functional. 

Isolated carboranethiols prefer the hollow site for dissociative adsorption on 
the gold (111) surface by forming strong Au-S bonds. PBE calculations indicate 
that M4 has the lowest total energy among the other isomers in the gas phase.
M1 has a dipole moment almost parallel to the surface while M9
has a moment nominally along the surface normal. Their 1 ML results indicate the
role of the moment orientation on the relative total supercell energies of SAMs.
The total energy of Au(111) with a full monolayer of M9 is always lower
that that of M1 with both the PBE and the optB86b-vdW functionals. 
On the other hand, M1 is more preferable than M9 in terms of stability 
since it always gives relatively stronger binding in both isolated and 
monolayer structures due to more favorable dipole-dipole interactions in SAMs. 

The LDA tends to underestimate the adsorption energies because of improper 
description of the long-range correlations. PBE calculations favor a 
(5$\times$5) film phase between the two competing surface structures since the 
dissociative adsorption energy per CT appears to be positive on 
($\sqrt{19}\times\!\sqrt{19}$)R23.4$^\circ$ structure. The inclusion of vdW 
interactions correct the average adsorption energy per CT of the latter phase 
to be slightly negative. Nevertheless, vdW-DF calculations still energetically 
prefer the (5$\times$5) geometry. Therefore, the vdW corrections are 
important to get reasonable results for carboranethiol-metal systems.

\section{Acknowledgements}
This work was financially supported by TUB\.{I}TAK 
under grant no. 213M182. EM and GG also acknowledge partial
support from Bal{\i}kesir University through BAP 2015/194.


\begin{thebibliography}{99}

\bibitem{Frasconi} Frasconi, M.; Mazzei, F.; Ferri, T. Protein immobilization 
at gold-thiol surfaces and potential for biosensing. \textit{Anal. Bioanal. 
Chem.} \textbf{2010}, 398, 1545-1564.

\bibitem{Love} Love, J. C.; Estroff, L. A.; Kriebel, J. K.; Nuzzo, R. G.; 
Whitesides, G. M. Self-assembled monolayers of thiolates on metals as a form of 
nanotechnology. \textit{Chem. Rev.} \textbf{2005}, 105, 1103-1169.

\bibitem{Schreiber} Schreiber, F. Structure and growth of self-assembling 
monolayers. \textit{Prog. Surf. Sci.} \textbf{2000}, 65, 151-256.

\bibitem{Vericat} Vericat, C.; Vela, M. E.; Benitez, G.; Carro, P.; Salvarezza, 
R. C. Self-assembled monolayers of thiols and dithiols on gold: new challenges 
for a well-known system. \textit{Chem. Soc. Rev.} \textbf{2010}, 39, 1805-1834.

\bibitem{Albayrak1} Albayrak, E.; Karabuga, S.; Bracco, G.; Danisman, M. F.  
Investigation of the deposition and thermal behavior of striped phases of 
unsymmetric disulfide self-assembled monolayers on Au (111): The case of 
11-hydroxyundecyl decyl disulfide. \textit{J. Chem. Phys.} \textbf{2015}, 142, 
014703.

\bibitem{Albayrak2} Albayrak, E.; Karabuga, S.; Bracco, G.; Danisman, M. F. 
11-Hydroxyundecyl octadecyl disulfide self-assembled monolayers on Au(111).
\textit{Appl. Surf. Sci.} \textbf{2014}, 311, 643-647.

\bibitem{Albayrak3} Albayrak, E.; Danisman, M. F. Helium Diffraction Study of 
Low Coverage Phases of Mercaptoundecanol and Octadecanethiol Self-Assembled 
Monolayers on Au(111) Prepared by Supersonic Molecular Beam Deposition.
\textit{J. Phys. Chem. C} \textbf{2013}, 117, 9801-9811.

\bibitem{Gronbeck1} Gr\"{o}nbeck, H.; Curioni, A.; Andreoni, W.  
Thiols and Disulfides on the Au(111) Surface: The Headgroup-Gold Interaction.
\textit{J. Am. Chem. Soc.} \textbf{2000}, 122, 3839-3842.

\bibitem{Ferrighi} Ferrighi, L.; Pan, Y.-X.; Gr\"{o}nbeck H.; Hammer, B. 
Study of Alkylthiolate Self-assembled Monolayers on Au(111) Using a Semilocal 
meta-GGA Density Functional. \textit{J. Phys. Chem. C} \textbf{2012}, 116, 
7374-7379.

\bibitem{Hayashi} Hayashi, T.; Morikawa, Y.; Nozoye, H. Adsorption state of 
dimethyl disulfide on Au(111): Evidence for adsorption as thiolate at the 
bridge site. \textit{J. Chem. Phys.} \textbf{2001}, 114, 7615-7621.

\bibitem{Gronbeck2} Gr\"{o}nbeck, H.; H\"{a}kkinen, H.; Whetten, R. L.
Gold-thiolate complexes form a unique c(4$\times$2) structure on Au(111).
\textit{J. Phys. Chem. C} \textbf{2008}, 112, 15490-15492.

\bibitem{Gronbeck3} Gr\"{o}nbeck, H.; H\"{a}kkinen, H. Polymerization at the 
Alkylthiolate-Au(111) Interface. \textit{J. Phys. Chem. B} \textbf{2007}, 111, 
3325-3327.

\bibitem{Yourdshahyan1} Yourdshahyan, Y.; Rappe, A. M. Structure and energetics 
of alkanethiol adsorption on the Au(111) surface. \textit{J. Chem. Phys.} 
\textbf{2002}, 117, 825-833.

\bibitem{Yourdshahyan2} Yourdshahyan, Y.; Zhang, H. K.; Rappe, A. M.  
$n$-alkyl thiol head-group interactions with the Au(111) surface. \textit{Phys. 
Rev. B} \textbf{2001}, 63, 081405(R).

\bibitem{Vargas} Vargas, M. C.; Giannozzi, P.; Selloni, A.; Scoles, G.  
Coverage-Dependent Adsorption of CH$_3$S and (CH$_3$S)$_2$ on Au(111): a 
Density Functional Theory Study. \textit{J. Phys. Chem. B} \textbf{2001}, 105, 
9509-9513.

\bibitem{Molina} Molina, L. M.; Hammer, B. Theoretical study of thiol-induced 
reconstructions on the Au(111) surface. \textit{Chem. Phys. Lett.} 
\textbf{2002}, 360, 264-271.

\bibitem{Otalvaro} Ot\'{a}lvaro, D.; Veening, T.; Brocks, G. Self-Assembled 
Monolayer Induced Au(111) and Ag(111) Reconstructions: Work Functions and 
Interface Dipole Formation. \textit{J. Phys. Chem. C} \textbf{2012}, 116 
7826-7837.

\bibitem{Rusu} Rusu, P. C.; Brocks, G. Surface Dipoles and Work Functions of 
Alkylthiolates and Fluorinated Alkylthiolates on Au(111). \textit{J. Phys. Chem. 
B} \textbf{2006}, 110, 22628-22634.

\bibitem{Fertitta} Fertitta, E.; Voloshina, E.; Paulus, B. Adsorption of 
multivalent alkylthiols on Au(111) surface: insights from DFT.
\textit{J. Comput. Chem.} \textbf{2014}, 35, 204-213.

\bibitem{AbuHusein} Abu-Husein, T.; Schuster, S.; Egger, D. A.; Kind, M.; 
Santowski, T.; Wiesner, A.; Chiechi, R.; Zojer, E.; Terfort, A.; Zharnikov, M.
The Effects of Embedded Dipoles in Aromatic Self-Assembled Monolayers.
\textit{Adv. Funct. Mat.} \textbf{2015}, 25, 3943-3957.

\bibitem{Verwuester} Verwuester, E.; Hofmann, O. T.; Egger, D. A.; Zojer, E. 
Electronic Properties of Biphenylthiolates on Au(111): The Impact of Coverage 
Revisited. \textit{J. Phys. Chem. C} \textbf{2015}, 119, 7817-7825. 

\bibitem{Fajin} Fajin, J. L. C.; Teixeira, F.; Gomes, J. R. B.; Cordeiro, M. N. D. S.  
Effect of van der Waals interactions in the DFT description of self-assembled 
monolayers of thiols on gold. \textit{Theo. Chem. Acc.} \textbf{2015}, 134, 1-13. 

\bibitem{Lustemberg} Lustemberg, P. G.; Abufager, P. N.; Martiarena, M. L.;  
Busnengo, H. F. Adsorption of methanethiol on Au(1 1 1): Role of hydrogen bonds. 
\textit{Chem. Phys. Lett.} \textbf{2014}, 610, 381-387.

\bibitem{YWang} Wang, Y.; Solano Canchaya, J. G.; Dong, W.; Alcami, M.;  
Busnengo, H. F.; Martin, F. Chain-Length and Temperature Dependence of 
Self-Assembled Monolayers of Alkylthiolates on Au(111) and Ag(111) Surfaces.
\textit{J. Phys. Chem. A} \textbf{2014}, 118, 4138-4146.

\bibitem{Torres1} Torres, E.; Blumenau, A. T.; Biedermann, P. U.  Steric and 
chain length effects in the ($\sqrt{3}\!\times\!\!\sqrt{3}$)R30$^\circ$ structures 
of alkanethiol self-assembled monolayers on Au(111). \textit{Chemphyschem} 
\textbf{2011}, 12, 999-1009.

\bibitem{Torres2} Torres, E.; Blumenau, A. T.; Biedermann, P. U. Mechanism 
for phase transitions and vacancy island formation in alkylthiol/Au(111) 
self-assembled monolayers based on adatom and vacancy-induced reconstructions.
\textit{Phys. Rev. B} \textbf{2009}, 79, 075440.

\bibitem{Longo} Longo, G. S.; Bhattacharya, S. K.; Scandolo, S. A molecular 
dynamics study of the role of adatoms in SAMs of methylthiolate on Au(111): 
A new force field parameterized from ab initio calculations. 
\textit{J. Phys. Chem. C} \textbf{2012}, 116, 14883-14891.

\bibitem{Cossaro} Cossaro, A.; Mazzarello, R.; Rousseau, R.; Casalis, L.; 
Verdini, A.; Kohlmeyer, A.; Floreano, L.; Scandolo, S.; Morgante, A.; Klein, M. 
L.; Scoles, G. X-ray Diffraction and Computation Yield the Structure of 
Alkanethiols on Gold(111). \textit{Science} \textbf{2008}, 321, 943-946.

\bibitem{Mazzarello} Mazzarello, R.; Cossaro, A.; Verdini, A.; Rousseau, R.; 
Casalis, L.; Danisman, M. F.; Floreano, L.; Scandolo, S.; Morgante, A.; Scoles, 
G. Structure of a CH$_3$S Monolayer on Au(111) Solved by the Interplay between 
Molecular Dynamics Calculations and Diffraction Measurements. \textit{Phys. 
Rev. Lett.} \textbf{2007}, 98, 016102.

\bibitem{Rousseau} Rousseau, R.; De Renzi, V.; Mazzarello, R.; Marchetto, D.;  
Biagi, R.; Scandolo, S.; del Pennino, U. Interfacial Electrostatics of 
Self-Assembled Monolayers of Alkane Thiolates on Au(111): Work Function 
Modification and Molecular Level Alignments. \textit{J. Phys. Chem. B} 
\textbf{2006}, 110, 10862-10872.

\bibitem{DeRenzi1} De Renzi, V.; Rousseau, R.; Marchetto, D.; Biagi, R.; 
Scandolo S.; del Pennino, U. Metal Work-Function Changes Induced by Organic 
Adsorbates: A Combined Experimental and Theoretical Study.
\textit{Phys. Rev. Lett.} \textbf{2005}, 95, 046804.

\bibitem{ZZhang} Zhang, L. Z.; Goddard, W. A.; Jiang, S. Y. Molecular 
simulation study of the c(4$\times$2) superlattice structure of alkanethiol 
self-assembled monolayers on Au(111). \textit{J. Chem. Phys.} \textbf{2002}, 
117, 7342-7349.

\bibitem{Fischer} Fischer, D.; Curioni, A.; Andreoni, W. Decanethiols on 
Gold: The Structure of Self-Assembled Monolayers Unraveled with Computer 
Simulations. \textit{Langmuir} \textbf{2003}, 19, 3567-3571.

\bibitem{Osella} Osella, S.; Cornil, D.; Cornil, J. Work function modification 
of the (111) gold surface covered by long alkanethiol-based self-assembled 
monolayers. \textit{Phys. Chem. Chem. Phys.} \textbf{2014}, 16, 2866-2873.

\bibitem{Barmparis} Barmparis, G. D.; Honkala, K.; Remediakis, I. N.  
Thiolate adsorption on Au(hkl) and equilibrium shape of large 
thiolate-covered gold nanoparticles. \textit{J. Chem. Phys.} \textbf{2013}, 138, 
064702.

\bibitem{JGWang1} Wang, J.-G.; Selloni, A. Influence of End Group and 
Surface Structure on the Current-Voltage Characteristics of Alkanethiol 
Monolayers on Au(111). \textit{J. Phys. Chem. A} \textbf{2007}, 111, 
12381-12385.

\bibitem{JGWang2} Wang, J.-G.; Selloni, A. The c(4$\times$2) Structure of 
Short- and Intermediate-Chain Length Alkanethiolate Monolayers on Au(111):
A DFT Study. \textit{J. Phys. Chem. C} \textbf{2007}, 111, 12149-12151.

\bibitem{Sun} Sun, Q.; Selloni, A. Interface and Molecular Electronic Structure 
vs Tunneling Characteristics of CH$_3$- and CF$_3$-Terminated Thiol Monolayers 
on Au(111). \textit{J. Phys. Chem. A} \textbf{2006}, 110, 11396-11400.

\bibitem{DeRenzi2} De Renzi, V.; Di Felice, R.; Marchetto, D.; Biagi, R.; del 
Pennino U.; Selloni, A. Ordered (3$\times$4) High-Density Phase of 
Methylthiolate on Au(111). \textit{J. Phys. Chem. B} \textbf{2004}, 108, 
16-20.

\bibitem{Heimel} Heimel, G.; Romaner, L.; Zojer, E.; Bredas, J.-L.  
The Interface Energetics of Self-Assembled Monolayers on Metals.
\textit{Acc. Chem. Res.} \textbf{2008}, 41, 721-729.

\bibitem{Campbell} Campbell, I. H.; Rubin, S.; Zawodzinski, T. A.; Kress, J. 
D.; Martin, R. L.; Smith, D. L.; Barashkov, N. N.; Ferraris, J. P.  
Controlling Schottky energy barriers in organic electronic devices using 
self-assembled monolayers. \textit{Phys. Rev. B} \textbf{1996}, 54, 14321-14324.

\bibitem{Chen} Chen, C.-Y.; Wu, K.-Y.; Chao, Y.-C.; Zan, H.-W.; Meng, H.-F.;  
Tao, Y.-T. Concomitant tuning of metal work function and wetting property with 
mixed self-assembled monolayers. \textit{Org. Elect.} \textbf{2011}, 12, 
148-153.

\bibitem{Venkataraman} Venkataraman, N. V.; Zuercher, S.; Rossi, A.; Lee, S.;  
Naujoks N.; Spencer, N. D. Spatial Tuning of the Metal Work Function by Means of 
Alkanethiol and Fluorinated Alkanethiol Gradients. \textit{J. Phys. 
Chem. C} \textbf{2009}, 113, 5620-5628.

\bibitem{Alloway} Alloway, D. M.; Graham, A. L.; Yang, X.; Mudalige, A.; 
ColoradoJr., R.; Wysocki, V. H.; Pemberton, J. E.; Lee, T. R.; Wysocki, R. J.;  
Armstrong, N. R. Tuning the Effective Work Function of Gold and Silver Using 
$\omega$-Functionalized Alkanethiols: Varying Surface Composition through 
Dilution and Choice of Terminal Groups. \textit{J. Phys. Chem. C} 
\textbf{2009}, 113, 20328-20334.

\bibitem{Wu} Wu, K.-Y.; Yu , S.-Y.; Tao, Y.-T. Continuous Modulation of 
Electrode Work Function with Mixed Self-Assembled Monolayers and Its Effect in 
Charge Injection. \textit{Langmuir} \textbf{2009}, 25, 6232-6238.

\bibitem{Xu} Xu, Y.; Baeg, K.-J.; Park, W.-T.; Cho, A.; Choi, E.-Y.; Noh, Y.-Y. 
Regulating Charge Injection in Ambipolar Organic Field-Effect Transistors 
by Mixed Self-Assembled Monolayers. \textit{ACS Appl. Mater. Interfaces} 
\textbf{2014}, 6, 14493-14499.

\bibitem{Lubben} Lubben, J. F.; Base, T.; Rupper, P.; Kunniger, T.; Machacek, 
J.; Guimond, S. Tuning the surface potential of Ag surfaces by chemisorption of 
oppositely-oriented thiolated carborane dipoles. \textit{J. Colloid Int. Sci.} 
\textbf{2011}, 354, 168-174.

\bibitem{Base1} Base, T.; Bastl, Z.; Havranek, V.; Lang, K.; Bould, J.; 
Londesborough, M. G. S.; Machacek, J.; Plesek, J. Carborane-thiol-silver 
interactions. A comparative study of the molecular protection of silver 
surfaces. \textit{Surf. Coat. Tech.} \textbf{2010}, 204, 2639-2646.

\bibitem{Base2} Base, T.; Bastl, Z.; Plzak, Z.; Grygar, T.; Plesek, J.; Carr, M. 
J.; Malina, V.; Subrt, J.; Bohacek, J.; Vecernikova, E.; Kriz, O.
Carboranethiol-Modified Gold Surfaces. A Study and Comparison of Modified 
Cluster and Flat Surfaces. \textit{Langmuir} \textbf{2005}, 21, 7776-7785.

\bibitem{Base3} Base, T.; Bastl, Z.; Slouf, M.; Klementova, M.; Subrt, J.; 
Vetushka, A.; Ledinsky, M.; Fejfar, A.; Machacek, J.; Carr, M. J.; 
Londesborough, M. G. S. Gold Micrometer Crystals Modified with Carboranethiol 
Derivatives. \textit{J. Phys. Chem. C} \textbf{2008}, 112, 14446-14455.

\bibitem{Scholz} Scholz, F.; Nothofer, H. G.; Wessels, J. M.; Nelles, G.; von 
Wrochem, F.; Roy, S.; Chen, X. D.; Michl, J. Permethylated 12-Vertex 
$p$-Carborane Self-Assembled Monolayers. \textit{J. Phys. Chem. C} 
\textbf{2011}, 115, 22998-23007.

\bibitem{Wrochem} von Wrochem, F.; Scholz, F.; Gao, D. Q.; Nothofer, H. G.; 
Yasuda, A.; Wessels, J. M.; Roy, S.; Chen, X. D.; Michl, J. 
High-Band-Gap Polycrystalline Monolayers of a 12-Vertex $p$-Carborane on 
Au(111). \textit{J. Phys. Chem. Lett.} \textbf{2010}, 1, 3471-3477.

\bibitem{Kim} Kim, J.; Rim, Y. S.; Liu, Y.; Serino, A. C.; Thomas, J. C.; Chen, 
H.; Yang, Y.; Weiss, P. S. Interface Control in Organic Electronics Using Mixed 
Monolayers of Carboranethiol Isomers. \textit{Nano Lett.} \textbf{2014}, 14, 
2946-2951.

\bibitem{Hohman1} Hohman, J. N.; Claridge, S. A.; Kim, M.; Weiss, P. S.  
Cage molecules for self-assembly. \textit{Mater. Sci. Eng. R-Rep.} 
\textbf{2010}, 70, 188-208.

\bibitem{Hohman2} Hohman, J. N.; Zhang, P.; Morin, E. I.; Han, P.; Kim, M.; 
Kurland, A. R.; McClanahan, P. D.; Balema V. P.; Weiss, P. S. Self-Assembly of 
Carboranethiol Isomers on Au{111}: Intermolecular Interactions Determined by 
Molecular Dipole Orientations. \textit{ACS Nano} \textbf{2009}, 3, 527-536.

\bibitem{Bould} Bould, J.; Machacek, J.; Londesborough, M. G. S.; Macias, R.;  
Kennedy, J. D.; Bastl, Z.; Rupper, P.; Base, T. Decaborane Thiols as Building 
Blocks for Self-Assembled Monolayers on Metal Surfaces. \textit{Inorg. Chem.} 
\textbf{2012}, 51, 1685-1694.

\bibitem{Thomas1} Thomas, J. C.; Schwartz, J. J.; Hohman, J. N.; Claridge, S. 
A.; Auluck, H. S.; Serino, A. C.; Spokoyny, A. M.; Tran, G.; Kelly, K. F.; 
Mirkin, C. A.; Gilles, J.; Osher, S. J.; Weiss, P. S. Defect-Tolerant Aligned 
Dipoles within Two-Dimensional Plastic Lattices. \textit{ACS Nano} 
\textbf{2015}, 9, 4734-4742.

\bibitem{Thomas2} Thomas, J. C.; Boldog, I.; Auluck, H. S.; Bereciartua, P. J.; 
Dusek, M.; Machacek, J.; Bastl, Z.; Weiss, P. S.; Base, T. Self-Assembled 
$p$-Carborane Analogue of $p$-Mercaptobenzoic Acid on Au\{111\}.
\textit{Chem. Mater.} \textbf{2015}, 27, 5425-5435.


\bibitem{Blochl} Bl\"{o}chl, P. E. Projector Augmented-Wave Method.
\textit{Phys. Rev. B} \textbf{1994}, 50, 17953. 

\bibitem{Kresse1} Kresse, G.; Hafner, J. Ab initio Molecular Dynamics for 
Liquid Metals. \textit{Phys. Rev. B} \textbf{1993}, 47, 558.

\bibitem{Kresse2} Kresse, G.; Furthm\"{u}ller, J. Efficient Iterative Schemes 
for Ab Initio Total-Energy Calculations Using a Plane-Wave Basis Set.
\textit{Phys. Rev. B} \textbf{1996}, 54, 11169.

\bibitem{Kresse3} Kresse, G.; Joubert, J. From Ultrasoft Pseudopotentials to 
the Projector Augmented-Wave Method. \textit{Phys. Rev. B} 
\textbf{1999}, 59, 1758.

\bibitem{Klimes} Klime\v{s}, J.; Bowler, D. R.; Michaelides, A. Van der Waals 
Density Functionals Applied to Solids. \textit{Phys. Rev. B} 
\textbf{2011}, 83, 195131.

\bibitem{Dion} Dion, M.; Rydberg, H.; Schr\"{o}der, E.; Langreth, D. C.; 
Lundqvist, B. I. Van der Waals Density Functional for General Geometries.
\textit{Phys. Rev. Lett.} \textbf{2004}, 92, 246401.

\bibitem{Wyckhoff} Wyckhoff, R. G. ``Crystal Structures", 2nd ed., 
Interscience Publishers, New York, 1958.

\end{thebibliography}
\end{document}